\documentclass[a4paper,11pt]{article}
\title{\bf Quantizing with a Higher Time Derivative}
\author{Sergei V. Ketov~${}^{a,b}$, Genta Michiaki~${}^{a}$ 
and Tsukasa Yumibayashi~${}^{a}$}
\date{}

\usepackage{amsmath,amsthm,amssymb}
\usepackage{mathrsfs}
\usepackage{graphicx}
\usepackage{booktabs}
\usepackage{bm}

\numberwithin{equation}{section}

\def\la#1{\label{eq: #1}} 
\def\re#1{\ref{eq: #1}} 

\def\bra#1{\left< #1 \right|}
\def\ket#1{\left| #1 \right>}

\def\eq#1{\begin{equation}#1\end{equation}}

\begin{document}
\maketitle

\begin{center}
${}^a$ {\it Department of Physics, Graduate School of Science, 
   Tokyo Metropolitan University, Hachioji-shi, Tokyo 192-0397, Japan}\\
${}^b$ {\it Institute for Physics and Mathematics of Universe, The 
           University of Tokyo, Kashiwa-shi, Chiba 277-8568, Japan}
\vglue.1in
ketov@phys.se.tmu.ac.jp, michiaki-genta@ed.tmu.ac.jp, 
yumibayashi-tsukasa@ed.tmu.ac.jp

\vglue.2in

{\bf Abstract}
\vglue.1in

\end{center}

\noindent We review the classical and quantum theory of the Pais-Uhlenbeck oscillator 
as the toy-model for quantizing $f(R)$ gravity theories. 
\vglue.2in

\section{Introduction}

It is commonplace in {\it Quantum Field Theory} (QFT) that a QFT with higher 
(time) derivatives is believed to be doomed from the point of view of physics, because 
of ghosts or states of negative norm, and thus it should be dismissed. The standard 
reference is 
the very old result (known in the literature as the Ostrogradski theorem \cite{ost}) 
claiming a linear instability in any Hamiltonian system associated with the Lagrangian 
having the higher (ie. more than one) time derivative that cannot be eliminated
by partial integration.  

The key point of the Ostrogradski method \cite{ost} is a canonical quantization of
the clasically equivalent theory without higher derivatives via considering the higher 
derivatives of the initial coordinates as the {\it independent} variables. 

The interest in the higher-derivative QFT was recently revived due to some
novel developments in the gravitational theory, related to the so-called
$f(R)$-gravity theories -- see eg., ref.~\cite{tsu} for a review. The $f(R)$
gravity theories are defined by replacing the scalar curvature $R$ in the 
Einstein action by a function $f(R)$. The $f(R)$ gravity theories give 
the self-consistent non-trivial alternative to the standard $\Lambda$-CDM Model
of Cosmology, by providing the geometrical phenomenological description of inflation
in the early universe and Dark Energy in the present universe. Despite of 
the apparent presence of the higher derivatives, a classical $f(R)$ gravity theory 
can be free of ghosts and tachyons. A supersymmetric extension of $f(R)$ gravity was 
recently constructed in superspace \cite{our1}.

Already the simplest model of $(R+R^2)$ gravity \cite{star} is known as the viable
model of chaotic inflation, because it is consistent with the recent WMAP measurements of
the Cosmic Macrowave Background (CMB) radiation \cite{wmap}. Its supersymmetric extension
was recently constructed in refs.~\cite{kstar,2kw}.

On the one side, any quadratically generated (with respect to the curvature) quantum 
theory of gravity has ghosts in its perturbative quantum propagator \cite{haw}. However,
on the other side, any $f(R)$ gravity theory is known to be {\it classically} equivalent 
to the scalar-tensor gravity (ie. to the usual quintessence) \cite{eq,bac,ma}, while the
stability conditions in the $f(R)$ gravity ensure the ghost-and-tachyon-freedoom of the
classically equivalent quintessence theory \cite{kkw,kw1}. It now appears that in some 
cases the presence of the higher derivatives may be harmless \cite{wood}.  It also gives 
rise to the non-trivial natural question of how to make sense out of the quantized 
$f(R)$ gravity?  

The $f(R)$ gravity theories are just the particular case of the higher-derivative quantum 
gravity theories which have been investigated in the past. They were found to be 
renormalizable
\cite{stelle} and asymptotically free \cite{frat}.  A generic higher-derivative gravity
suffers, however, from the presence of ghosts and states of negative norm which apparently
spoil those QFT from physical applications. However, the issue of ghosts and their
physical interpretation deserves a more detailed study. The complexity of the  
higher-derivative gravity is the formidable technical obstacle for that. It is, therefore,
of interest to consider simpler QFT as the toy-models.
 
Similar features (like renormalizability and asymptotic freedom) exhibit the quantum 
Non-Linear Sigma-Models with higher derivatives, which have striking similarities to the
higher-derivative quantum gravity  \cite{buchk,per,kbook}. However, even those QFT are 
too complicated because of their high degree of non-linearity. 

Perhaps, the simplest toy-model is given by the {\it Pais-Uhlenbeck} (PU) quantum
oscillator in Quantum Mechanics \cite{pu}. As was demonstrated by Hawking and Hertog 
\cite{hh}, it may be possible to give physical meaning to the Euclidean path integral of 
the PU oscillator, as the set of consistent rules for calculation of observables, even 
when ``living with ghosts''. The basic idea of ref.~\cite{hh} is to abandom unitarity,
while never producing and observing negative norm states. 

The idea of Hawking and Hertog found further support in refs.~\cite{zer,pol} where the 
physical propagator of the PU oscillator was calculated by using the van Vleck-Pauli 
approach (the saddle point method for the Euclidean path integral) and Forman's theorem 
\cite{forman}. In this paper we systematically review the classical and quantum theory of 
the PU oscillator from the first principles, along the lines of 
refs.~\cite{wood,hh,zer,pol}. 

Some mathematical connections between the higher-derivative particle models, higher spins and 
noncommutative geometry can be found in ref.~\cite{ply}.

\newpage

\section{Ostrogradski method with higher derivatives}

Consider a one-dimensional mechanical system with the action
\eq{
S[q] = \int dt \mspace{5mu} L \bigl( q,Dq,\cdots ,D^n q \bigr) \la{generalaction}
}
in terms of the Lagrange function $L$ of $q(t)$ and its time derivatives, 
where $n \geq 2$ and $D = \frac{d}{dt}$. The Euler-Lagrange equation reads
\eq{
\sum_{i=0}^n (-D)^i \frac{\partial L}{\partial (D^i q)} = 0 \la{ELeq}
}

The Ostrogradski method \cite{ost} gives the Hamiltonian formulation of the higher 
derivative Lagrange formulation by introducing more independent variables. 

The independent generalized coordinates $Q_i$ are defined by 
\eq{
Q_i = D^{i-1} q \mspace{50mu} \bigl( i = 1 , \cdots , n \bigr) \la{Qi1}
}
The generalized momentum $P_n$ is defined by
\eq{
\genfrac{.}{|}{}{}{\partial L}{\partial (D^nq)}_{ 
\genfrac{.}{.}{0pt}{}{ D^{i-1}q = Q_i }{ D^nq = A } } = P_n \la{Pn}
}
There are $n+1$ independent variables $\{ Q_1,\cdots ,Q_n,P_n \}$ that are in 
correspondence to the $n+1$ variables  $\{ D^0q,\cdots ,D^nq \}$ of the higher derivative 
action (\re{generalaction}).

By solving eq.(\re{Pn}) with respect to $A=D^n q$ (assuming that it is possible), one gets
\eq{
D^n q = A ( Q_1 , \cdots , Q_n , P_n )
}
Therefore, the Lagrange dynamics can be represented in terms of the $n+1$ independent 
variables 
$\{ Q_1,\cdots ,Q_n,P_n \}$ as
\eq{
L = L \bigl( Q_1 , \cdots , Q_n , A ( Q_1 , \cdots , Q_n , P_n ) \bigr)
}

A Legendre transformation is used to pass from the Lagrange formulation to 
the Hamiltonian one. With the generalized coordinates $\{ Q_1,\cdots ,Q_n \}$ and the 
generalized momentum $P_n$ as the independent variables, the total differential of the 
Lagrangian is given by
\begin{eqnarray}
dL &=& \sum_{j=1}^n \genfrac{.}{|}{}{}{\partial L}{\partial (D^{j-1}q)}_{ 
\genfrac{.}{.}{0pt}{}{ D^{i-1}q = Q_i }{ D^nq = A } } dQ_j
+ P_n dA \la{dL1} \nonumber \\
&=& \frac{\partial L}{\partial q} dQ_1 + \sum_{j=2}^n \frac{\partial L}{
\partial (D^{j-1}q)} dQ_j + P_n dA \nonumber \\
&=& D \sum_{j=1}^n (-D)^{j-1} \frac{\partial L}{\partial (D^j q)} dQ_1
+ \sum_{j=1}^{n-1} \frac{\partial L}{\partial (D^jq)} dQ_{j+1} + P_n dA \nonumber \\
& &
\end{eqnarray}
where we have used  eqs.~(\re{ELeq}) and (\re{Pn}), and
\eq{
dA = \sum_{j=1}^n \frac{\partial A}{\partial Q_j} dQ_j +
 \frac{\partial A}{\partial P_n} dP_n
}

Let us now define the $n-1$ generalized momenta as
\eq{
P_i = \sum_{j=i}^n (-D)^{j-i} \frac{\partial L}{\partial (D^j q)} \mspace{50mu} \bigl( 
i = 1 , \cdots , n-1 \bigr) \la{Pi1}
}
They satisfy the relations
\eq{
\frac{\partial L}{\partial (D^iq)} = P_i + D P_{i+1}
}
Therefore, eq.~(\re{dL1}) can be rewritten to the form
\eq{
d\biggl[ \sum_{i=1}^{n-1} P_i (DQ_i) + P_n A - L \biggr] = - \sum_{i=1}^{n} (DP_i) dQ_i 
+ \sum_{i=1}^{n} (DQ_i) dP_i \la{dH}
}

Equation (\re{dH}) gives rise to the Hamiltonian in the form
\eq{
H = \sum_{j=1}^{n-1} P_j (DQ_j) + P_n A - L \la{Hostro}
}
The Hamilton equations of motion are given by
\eq{
DQ_i = \frac{\partial H}{\partial P_i} \mspace{50mu} \text{and} \mspace{50mu} DP_i = - \frac{\partial H}{\partial Q_i}
}

\section{PU oscillator}

The PU oscillator \cite{pu} is an extension of the harmonic oscillator with the higher
time derivatives, and  is the particular case of the higher-derivative theory
introduced in Sec.~2. The special features of the PU opscilator are\\
(i) the equation of motion is {\it linear}:
\eq{
F(D)q = 0
}
where $F$ is a linear differential operator;\\
(ii) the $F$ is {\it polynomial} (with respect to $D$) with {\it constant} coefficients:
\eq{
F(D) = \sum_{i=0}^n a_i D^i
}
where $a_0,\cdots,a_n$ are the real constants;\\
(iii) there is the time reversal invariance with respect to $t\rightarrow -t$.
Hence, the polynomial $F$ has only even powers of the time derivative $D$.

The Lagrangian of the one-dimensional PU oscillator reads
\eq{
L \bigl( q, Dq ,\cdots ,D^nq \bigr) = - \sum_{i=0}^n \frac{a_i}{2} (D^i q)^2 \mspace{50mu}
\bigl( a_0 \neq 0, a_n \neq 0 \bigr) \la{generalLPU}
}
where $a_i \mspace{5mu} (i=0,\cdots ,n)$ are real constants. The Euler-Lagrange equation
of motion is given by
\begin{align}
0 =& \sum_{i=0}^n (-D)^i \biggl[ - a_i D^i q \biggr] \nonumber \\
=& - a_0 \biggl[ \sum_{i=0}^n (-1)^i \frac{a_i}{a_0} D^{2i} \biggr] q
\end{align}
Accordingly, the differential operator $F(D)$ reads 
\eq{
F(D) = \sum_{i=0}^n (-1)^i \frac{a_i}{a_0} D^{2i} \la{F1}
}
The equation of motion can be rewritten to the form
\eq{
 F(D) q = 0 \la{generalEOM}
}
The PU Lagrangian takes the form (up to a boundary term)
\eq{
\bar{L} = - \frac{a_0}{2} q F(D) q \la{barL1}
}

The differential operator $F(D)$  can be brought to the factorized form
\eq{
F(D) = \prod_{i=1}^n \biggl( 1 + \frac{D^2}{\omega _i^2} \biggr) \la{F2}
}
where the constants $\omega _i \mspace{5mu} (i=1,\cdots ,n)$ are the solutions (roots) 
of the equation $F(i\omega)=0$. Let us introduce $n$ new operators 
\eq{
G_i(D) = \prod_{ \genfrac{.}{.}{0pt}{}{j=1}{j \neq i} }^n \biggl( 
1 + \frac{D^2}{\omega _j^2} \biggl)
\mspace{50mu} (i=1,\cdots ,n) \la{G}
}
and define the $n$ generalized coordinates as
\eq{
Q_i = G_i(D) q \mspace{50mu} (i=1,\cdots ,n) \la{Qi2}
}
Those generalized coordinates $Q_j$ are called {\it harmonic} coordinates.
By using the harmonic coordinates, the PU Euler-Lagrange eq.~(\re{generalEOM}) can be
 rewitten to the $n$ equations
\eq{
 \biggl[ 1 + \frac{D^2}{\omega _i^2} \biggr] Q_i = 0
}
It means that the PU oscillator can be interpreted as $n$ harmonic oscillators.
Accordingly, the PU Lagrangian (\re{barL1}) can be rewritten to the form
\eq{
\bar{L} = - \frac{a_0}{2} \sum_{i=1}^n \eta _i Q_i \biggl( 1 + 
\frac{D^2}{\omega _i^2} \bigg) Q_i \la{barL2}
}
where the $n$ constants $\eta _i$ have been introduced as
\eq{
\eta _i = \biggl( \omega _i^2 \genfrac{.}{|}{}{}{dF}{d(D^2)}_{D^2 = 
- \omega _i^2} \biggr) ^{-1} \la{etai}
}

To prove eq.~(\re{etai}), we first notice that it amounts to
\eq{
\sum_{i=1}^n \eta _i G_i(D) = 1
}
By the definiton of $G(D)$ in eq.(\re{G}) we have
\begin{eqnarray}
G_i(D^2=-\omega _j^2)
&=& \prod_{ \genfrac{.}{.}{0pt}{}{k=1}{k \neq i} }^n \biggl( 1 - \frac{\omega _j^2}{
\omega _k^2} \biggr) \nonumber \\
&=& \delta _{ij} \prod_{ \genfrac{.}{.}{0pt}{}{k=1}{k \neq j} }^n \biggl( 1 - \frac{
\omega _j^2}{\omega _k^2} \biggr)
\end{eqnarray}
so that
\begin{eqnarray}
\sum_{i=1}^n \eta _i G_i(D) &=& 1 \\
\sum_{i=1}^n \eta _i G_i(D^2=-\omega _j^2)
&=& \eta _j \prod_{ \genfrac{.}{.}{0pt}{}{k=1}{k \neq j} }^n \biggl( 1 
- \frac{\omega _j^2}{\omega _k^2} \biggr) = 1
\end{eqnarray}
indeed. Therefore, the constants $\eta _i$ are given by
\eq{
\eta _i = \biggl[ \prod_{ \genfrac{.}{.}{0pt}{}{k=1}{k \neq i} }^n
\bigg( 1 - \frac{\omega _i^2}{\omega _k^2} \biggr) \biggr]^{-1} \la{hodai1}
}
Next, we prove that
\eq{
\omega _i^2 \genfrac{.}{|}{}{}{dF}{dD^2}_{D^2 = - \omega _i^2} = \prod_{ 
\genfrac{.}{.}{0pt}{}{k=1}{k \neq i} }^n
\biggl( 1 - \frac{\omega _i^2}{\omega _k^2} \biggr) \la{hodai2}
}
By the use of eq.(\re{F2}) we find
\begin{eqnarray}
\frac{dF}{dD^2} &=& \frac{d}{dD^2} \prod_{j=1}^n \bigl( 1 + \frac{D^2}{\omega _j^2} 
\bigr) \nonumber \\
&=& \sum_{k=1}^n \frac{1}{\omega _k^2} \prod_{ \genfrac{.}{.}{0pt}{}{j=1}{j \neq k} }^n
\bigl( 1 + \frac{D^2}{\omega _j^2} \bigr) \nonumber \\
&=& \sum_{j=1}^n \frac{1}{\omega _j^2} G_j(D)
\end{eqnarray}
so that
\begin{eqnarray}
\genfrac{.}{|}{}{}{dF}{dD^2}_{D^2=-\omega _i^2} &=& \sum_{j=1}^n \frac{1}{\omega _j^2} 
G_j(D^2=-\omega _i^2) \nonumber \\
&=& \sum_{j=1}^n \frac{1}{\omega _j^2} \delta _{ji} \prod_{ \genfrac{.}{.}{0pt}{}{k=1}{k 
\neq i} }^n
\bigl( 1 - \frac{\omega _i^2}{\omega _k^2} \bigr) \nonumber \\
&=& \frac{1}{\omega _i^2} \prod_{ \genfrac{.}{.}{0pt}{}{k=1}{k \neq i} }^n
\bigl( 1 - \frac{\omega _i^2}{\omega _k^2} \bigr)
\end{eqnarray}
Equation (\re{hodai2}) is now confirmed and, hence, via eq.~(\re{hodai1}) also 
eq.~(\re{etai}) follows.

In terms of the harmonic coordinates (\re{Qi2}), the Lagrangian $\bar{L}$,
\begin{eqnarray}
\bar{L} &=& - \frac{a_0}{2} q F(D) q \nonumber \\
&=& - \frac{a_0}{2} \sum_{i=1}^n \eta _i Q_i \bigl( 1 + \frac{D^2}{\omega _i^2} \bigr) 
Q_i \la{barL3}
\end{eqnarray}
with the constants $\eta _i$ given by eq.~(\re{etai}), can be rewritten to the form
\eq{
\tilde{L} = \frac{a_0}{2} \sum_{i=1}^n \eta _i \bigl( \frac{1}{\omega _i^2} (DQ_i)^2 
- Q_i^2 \bigr) \la{tildeL}
}
up to a boundary term.

The Lagrangian (\re{tildeL}) is just a sum of the Lagrangians of $n$ harmonic 
oscillators. Hence, similarly to a free system of $n$ particles, we can change the 
Lagrangian formulation into the Hamiltonian formulation. We define the generalized 
momenta $P_i$ by taking the harmonic coordinates $Q_i$ and the velocities $DQ_i$ as
the Lagrange variables,
\begin{eqnarray}
P_i &=& \frac{\partial \tilde{L}}{\partial (DQ_i)} \nonumber \\
&=& \frac{a_0 \eta _i}{\omega _i^2} DQ_i \mspace{50mu} (i=1,\cdots ,n) \la{Pi2}
\end{eqnarray}

The system of $n$ free particles does not have higher derivatives, so its Hamiltonian is
\eq{
H = \sum_{i=1}^n P_i (DQ_i) - L
}
Equations (\re{tildeL}) and (\re{Pi2}) imply
\eq{
H = \sum_{i=1}^n \biggl( \frac{\omega _i^2}{2 a_0 \eta _i} P_i^2 + \frac{a_0 \eta _i}{2} 
Q_i^2 \biggr)
}

By rescaling the harmonic coordinates and the generalized momenta as
\eq{
Q_i \rightarrow \tilde{Q}_i = \frac{ \sqrt{a_0 |\eta _i|} }{ \omega _i } Q_i 
\mspace{50mu} \text{and} \mspace{50mu}
P_i \rightarrow \tilde{P}_i = \frac{ \omega _i \sqrt{|\eta _i|} }{ 
\eta _i \sqrt{a_0} } P_i
}
we get the final Hamiltonian
\eq{
H = \frac{1}{2} \sum_{i=1}^n \frac{\eta _i}{|\eta _i|} \bigl( \tilde{P_i}^2 + 
\omega _i^2 \tilde{Q_i}^2 \bigr)
}
The presence of both positive and negative values of the constants $\eta _i$ 
in the Hamiltonian implies both positive and negative values of energy. The 
constants $\eta _i$ are given by eq.~(\re{hodai1}). If $\omega _i$ satisfy 
$i<j \Rightarrow \omega _i<\omega _j$, the constants $\eta _i$ are positive for the odd 
number $i$, and are negative for the even number $i$. Therefore, the Hamiltonian is
\eq{
H = \frac{1}{2} \sum_{i=1}^n (-1)^{i-1} \biggl( \tilde{P_i}^2 + 
\omega _i^2 \tilde{Q_i}^2 \biggr)
}
This Hamiltonian can be interpreted as that of $n$ harmonic oscillators, with 
the positive and negative energy levels appearing alternatively. Because of that reason, 
the PU oscillator has an instability (for any interaction). It is related to a possible ghost 
state of negative norm in PU quantum theory (see Sec.~6). In what follows we consider 
the simplest case of PU oscillator with $n=2$ only.

\section{PU oscillator for $n=2$: explicit results}

Let us consider the Lagrangian 
\eq{
L = \frac{1}{2} \biggl( \frac{dq}{dt} \biggr) ^2 - V(q) - \frac{\alpha ^2}{2} 
\biggl( \frac{d^2 q}{dt^2} \biggr)^2
\mspace{50mu} \bigl( \text{where } \alpha \neq 0 \bigr) \la{Lagrangian}
}
with a scalar potential $V(q)$. In the case of the PU oscillator, the potential $V(q)$ is
a quadratic function of $q$. Since the (mass) dimension of time is $-1$ (in the natural 
units
$\hbar=c=1$), the dimension of the Lagrangian $L$ is $1$, the dimension of $q$ is $-1/2$, 
and that of the constant $\alpha$ is $-1$.

Let the trajectory $q$ be a sum of the classical trajectory $q_{cl}$ and the displacement 
$\tilde{q}$, ie. $q = q_{cl} + \tilde{q}$, where the classical trajectory $q_{cl}$ is a 
solution to the {\it equation of motion} (EOM) with the boundary conditions \cite{hh}
\eq{
\mathscr{A} : \mspace{20mu} q(0) = q_0, \mspace{20mu} q(T) = q_T, \mspace{20mu} 
\dot{q}(0) = \dot{q}_0,
\mspace{20mu} \dot{q}(T) = \dot{q}_T \la{boundary}
}
where the dots above stand for the time derivatives.

With the boundary conditions (\re{boundary}), the boundary condition of $\tilde{q}$ is
\eq{
\tilde{\mathscr{A}} : \mspace{20mu} \tilde{q}(0) = 0, \mspace{20mu} \tilde{q}(T) = 0, 
\mspace{20mu} \dot{\tilde{q}}(0) = 0,
\mspace{20mu} \dot{\tilde{q}}(T) = 0
}

The action of $q_{cl}+\tilde{q}$ is given by
\eq{
S[ q_{cl} + \tilde{q} ] = S[q_{cl}] + \int_0^T dt
\biggl( \frac{1}{2} \dot{\tilde{q}}^2 - V ( q_{cl} + \tilde{q} ) 
+ V(q_{cl}) + \tilde{q}V'(q_{cl})
- \frac{\alpha ^2}{2} \ddot{\tilde{q}}^2 \biggr) \la{action}
}
where we have introduced the notation
\eq{
V'(q_{cl}) = \genfrac{.}{|}{}{}{dV}{dq}_{q = q_{cl}}
}

In eq.(\re{action}) the term $V(q_{cl} + \tilde{q}) - V(q_{cl}) - \tilde{q}V'(q_{cl})$
represents the {\it gap} between the full action $S[q]$ and the classical action $S[q_{cl}]$,
which generically depends on both the classical trajectory $q_{cl}$ and the displacement 
$\tilde{q}$. After expanding the scalar potential $V$ in Taylor series,
\eq{
V(q_{cl} + \tilde{q}) = V(q_{cl}) + \tilde{q} V'(q_{cl}) + \frac{1}{2!} \tilde{q}^2 V''(q_{cl})	
+ \cdots
}
we find that, when the second derivative $V''$ is constant, the gap
$V(q_{cl} + \tilde{q}) - V(q_{cl}) - \tilde{q}V'(q_{cl})$ does {\it not} depend on the classical 
trajectory $q_{cl}$. It is the case when the potential $V$ is a quadratic function of $q$,
like the PU oscillator.

In the path integral quantization (sec.~7), the gap between the full action and the classical 
action is a quantum effect. When the potential is a quadratic function (like that of the 
PU oscillator), that quantum effect does depend on $\tilde{q}$, but does not depend on the 
classical trajectory. In what follows, we only consider a quadratic function for the
scalar potential in the form 
\eq{
V(q) = \frac{m^2}{2} q^2
}
ie. the scalar potential of a harmonic oscillator with the mass $m>0$, The Lagrangian is given by
\eq{
L_{PU} = \frac{1}{2} \dot{q}^2 - \frac{m^2}{2} q^2 
- \frac{\alpha ^2}{2} \ddot{q}^2 \la{LagrangianPU}
}

The parameter $\alpha$ measures a contribution of the second derivative to the harmonic
oscillator. Therefore, we can expect the classical trajectory to behave just like that of 
the harmonic oscillator when $\alpha$ is small.

The Euler-Lagrange EOM of the Lagrangian (\re{LagrangianPU}) are given by eq.(\re{ELeq}),
\begin{eqnarray}
0 &=& \sum_{i=0}^2 (-D)^i \frac{\partial L}{\partial (D^i q)} \nonumber \\
&=& - m^2 q - \ddot{q} - \alpha ^2 \ddddot{q}
\end{eqnarray}
or, equivalently,
\eq{
 \biggl( m^2 + D^2 + \alpha ^2 D^4 \biggr) q = 0 \la{EOM}
}

It is not difficult to find clasical solutions to the EOM in eq.~(\re{EOM}).
When searching for the classical trajectory in the oscillatory form 
$q_{cl} = \exp (i\lambda t)$, the EOM reads
\eq{
 \biggl( m^2 - \lambda ^2 + \alpha ^2 \lambda ^4 \biggr) e^{i\lambda t} = 0
}
and, therefore, we have
\eq{
\lambda ^2 = \frac{ 1 \pm \sqrt{1-4\alpha ^2 m^2} }{2\alpha ^2} \la{lambda}
}
When $\lambda$ is real, the Lagrangian $L_(PU)$ is an extension of the harmonic oscillator
indeed. Hence, we need the condition
\eq{
0 < \alpha m < \frac{1}{2} \la{am}
}
It means that the Lagrangina $L_{PU}$ has the oscillating solution which is similar to the 
trajectory of the harmonic oscillator. A general solution reads
\eq{
q(t) = A_+ \cos\bigl( \lambda _+ t \bigr) + B_+ \sin\bigl( \lambda _+ t \bigr)
+ A_- \cos\bigl( \lambda _- t \bigr) + B_- \sin\bigl( \lambda _- t \bigr) \la{qcl}
}
where $A_+,B_+,A_-,B_-$ are the integration constants, and 
\eq{
\lambda _{\pm} = \sqrt{ \frac{ 1 \mp \sqrt{1-4\alpha ^2 m^2} }{2\alpha ^2} } \la{lpm}
}
The values of the constants $(A_+,B_+,A_-,B_-)$ are determined by the boundary conditions.

The Hamiltonian formulation for the Lagrangian (\re{LagrangianPU}) can be obtained by the
Ostrogradski method. The generalized coodinates and momenta are given in Sec.~2, ie.
\begin{eqnarray}
Q_1 &=& q \mspace{50mu} \text{and} \mspace{50mu}
P_1 = \frac{\partial L}{\partial \dot{q}} - D \frac{\partial L}{\partial \ddot{q}} \nonumber \\
Q_2 &=& \dot{q} \mspace{50mu} \text{and} \mspace{50mu} P_2 = \frac{\partial L}{\partial 
\ddot{q}} \la{QPostro}
\end{eqnarray}
which imply
\begin{eqnarray}
P_1 &=& \dot{q} + \alpha ^2 \dddot{q} \nonumber \\
P_2 &=& - \alpha ^2 \ddot{q} \la{Postro}
\end{eqnarray}

The Hamiltonian is given by eq.(\re{Hostro}). ie.
\begin{eqnarray}
H &=& P_1 (DQ_1) + P_2 A - L \nonumber \\
&=& P_1 Q_2 - \frac{1}{2\alpha ^2} P_2^2 - \frac{1}{2} Q_2^2 + \frac{m^2}{2} Q_1^2
\end{eqnarray}
or, equivalently, 
\eq{
H = \alpha ^2 \dot{q} \dddot{q} - \frac{\alpha ^2}{2} \ddot{q}^2 
+ \frac{1}{2} \dot{q}^2 + \frac{m^2}{2} q^2 \la{H11}
}

Since the Hamiltonian does not evolve with time, we can find the energy by substituting 
$q(t)$ of eq.~(\re{qcl}) at $t=0$ into eq.~(\re{H11}), as well as 
$q,\dot{q},\ddot{q}$ and $\dddot{q}$ at $t=0$, ie.
\begin{eqnarray}
q (0) &=& A_+ + A_- \nonumber \\
\dot{q} (0) &=& B_+ \lambda _+ + B_- \lambda _- \nonumber \\
\ddot{q} (0) &=& - A_+ \lambda _+^2 - A_- \lambda _-^2  \\
\dddot{q} (0) &=& - B_+ \lambda _+^3 - B_- \lambda _-^3 \nonumber
\end{eqnarray}
It is now straightforward to calculate the Hamiltonian (\re{H11}). We find
\begin{eqnarray}
H &=& \alpha ^2 \dot{q}(0) \dddot{q}(0) - \frac{\alpha ^2}{2} \ddot{q}(0)^2 
+ \frac{1}{2} \dot{q}(0)^2 + \frac{m^2}{2} q(0)^2  \\
& = & \frac{1}{2} \lambda _+^2 \sqrt{1 - 4\alpha ^2 m^2}(A_+^2 + B_+^2)
- \frac{1}{2} \lambda _-^2 \sqrt{1 - 4\alpha ^2 m^2}(A_-^2 + B_-^2)\nonumber   \la{H12}
\end{eqnarray}

To get the Hamiltonian formulation in the harmonic coordinates, we begin with the EOM 
in the form (\re{EOM}), whose differentioal operator $F(D)$ is defined by
\eq{
F(D) = 1 + \frac{D^2}{m^2} + \frac{\alpha ^2 D^4}{m^2}
}
It can be factorized as
\eq{
F(D) = \biggl( 1 + \frac{D^2}{\lambda _+^2} \biggr) \biggl( 1 + \frac{D^2}{\lambda _-^2} \biggr)
}
where $\lambda _\pm$ are given by eq.~(\re{lpm}). Therefore, the harmonic coodinates are given by
\eq{
Q_+ = \biggl( 1 + \frac{D^2}{\lambda _-^2} \biggr) q \mspace{50mu} \text{and} \mspace{50mu}
Q_- = \biggl( 1 + \frac{D^2}{\lambda _+^2} \biggr) q \la{harco}
}

The constants $\eta _i$ of eq.~(\re{etai}) can be computed as follows. We have 
\eq{
\frac{dF}{dD^2} = \frac{1}{m^2} + \frac{2\alpha ^2 D^2}{m^2}
}
so that 
\begin{eqnarray}
\eta _\pm &=& \biggl( \lambda _\pm ^2 \left. \frac{dF}{dD^2}\right|_{D^2 = - \lambda _\pm ^2} 
\biggr) ^{-1} \nonumber \\
&=& \biggl( \frac{\lambda _\pm ^2}{m^2} ( 1 - 2\alpha ^2 \lambda _\pm ^2) \biggr) ^{-1} 
\nonumber \\
&=& \biggl( \pm \frac{\lambda _\pm ^2}{m^2} \sqrt{1 - 4\alpha ^2 m^2} \biggr) ^{-1} \nonumber \\
&=& \pm \frac{m^2}{ \lambda _\pm ^2 \sqrt{1 - 4\alpha ^2 m^2} }
\end{eqnarray}
Therefore, the generalized momenta in eq.~(\re{Pi2}) are 
\begin{eqnarray}
P_\pm &=& \frac{m^2 \eta _\pm}{\lambda _\pm ^2} DQ_\pm \nonumber \\
&=& \pm \frac{m^4}{ \lambda _\pm ^4 \sqrt{1 - 4\alpha ^2 m^2} } DQ_\pm \la{Pharmonic}
\end{eqnarray}
and the Hamiltonian is given by
\begin{eqnarray}
H &=& \sum_{j=\pm} \biggl( \frac{\lambda _j^2}{2 m^2 \eta _j} P_j^2 
+ \frac{m^2 \eta _j}{2} Q_j^2 \biggr) \nonumber \\
&=& \sum_{j=\pm} j \frac{m^4}{ 2\lambda _j^4 \sqrt{1 - 4\alpha ^2 m^2} } \biggl( (DQ_j )^2 
+ \lambda _j Q_j^2 \biggr) \la{H21}
\end{eqnarray}
where we have substituted the classical solution (\re{qcl}).

The harmonic coodinates (\re{harco}) read
\begin{eqnarray}
Q_+ &=& A_+ \biggl( 1 - \frac{\lambda _+^2}{\lambda _-^2} \biggr) \cos\bigl( \lambda _+ t \bigr)
+ B_+ \biggl( 1 - \frac{\lambda _+^2}{\lambda _-^2} \biggr) \sin\bigl( \lambda _+ t \bigr) \\
Q_- &=& A_- \biggl( 1 - \frac{\lambda _-^2}{\lambda _+^2} \biggr) \cos\bigl( \lambda _- t \bigr)
+ B_- \biggl( 1 - \frac{\lambda _-^2}{\lambda _+^2} \biggr) \sin\bigl( \lambda _- t \bigr)
\end{eqnarray}
where 
\begin{eqnarray}
1 - \frac{\lambda _\pm ^2}{\lambda _\mp ^2}
&=& \lambda _\pm ^2 \biggl( \frac{1}{\lambda _\pm ^2} - \frac{1}{\lambda _\mp ^2} \biggr) 
\nonumber \\
&=& \pm \frac{\lambda _\pm ^2}{m^2} \sqrt{1 - 4\alpha ^2 m^2}
\end{eqnarray}
Hence, we find 
\eq{
Q_\pm = \pm \frac{\lambda _\pm ^2}{m^2} \sqrt{1 - 4\alpha ^2 m^2}
\biggl( A_\pm \cos\bigl( \lambda _\pm t \bigr) + B_\pm \sin\bigl( \lambda _\pm t \bigr) \biggr) 
\la{Qharmonic}
}
Substituting them into the Hamiltonian (\re{H21}), we get
\eq{
H = \frac{1}{2} \lambda _+^2 \sqrt{1 - 4\alpha ^2 m^2} (A_+^2 + B_+^2)
- \frac{1}{2} \lambda _-^2 \sqrt{1 - 4\alpha ^2 m^2} (A_-^2 + B_-^2) \la{H22}
}
Equations (\re{H12}) and (\re{H22}) are {\it the same.} Therefore, we conclude that the 
Hamiltonian formulation by the Ostrogradski method is consistent with the Hamiltonian 
formulation in the harmonic coordinates, as they should.

The integration constants $(A_+,B_+)$ correspond to the harmonic oscillator with positive energy,
while the integration constants $(A_-,B_-)$ correspond to the harmonic oscillator with negative 
energy.

\section{Boundary conditions and spectrum}

Going back to the Lagrangian (\re{LagrangianPU}), let us consider its action over a finite time period 
$T$,
\eq{
S[q] = \int_0^T dt \mspace{5mu} L_{PU}
}
with the trajectory $q$ being a sum of the classical trajectory $q_{cl}$ and the displacement 
$\tilde{q}$, $q = q_{cl} + \tilde{q}$. In quantum theory, the displacement $\tilde{q}$ is a quantum 
coordinate. The action can be rewritten as
\eq{
S[q] = S[q_{cl}] + S[\tilde{q}] - \int_0^T dt \mspace{5mu} \biggl( \ddot{q_{cl}} + m^2 q_{cl} 
+ \alpha ^2 \ddddot{q}_{cl} \biggr) \tilde{q}
+ \biggl[ \dot{q}_{cl} \tilde{q} - \alpha ^2 \ddot{q}_{cl} \dot{\tilde{q}} + \alpha ^2 \dddot{q}_{cl}
 \tilde{q} 
\biggr] _0^T \la{Sexpansion}
}
Here the first term is the action of the classical trajectory $q_{cl}$, and the second term is the 
action of the quantum part $\tilde{q}$. The integrand of the third term vanishes because the classical 
trajectory is a solution of the (Euler-Lagrange) EOM. The fourth term depends on the boundary. However,
if the boundary condition on $\tilde{q}$ is given by
\eq{
\tilde{\mathscr{A}} : \mspace{20mu} \tilde{q}(0) = 0, \mspace{20mu} \tilde{q}(T) = 0,
\mspace{20mu} \dot{\tilde{q}}(0) = 0, \mspace{20mu} \dot{\tilde{q}}(T) = 0 \la{bcon}
}
the fourth term in eq.~(\re{Sexpansion}) also vanishes. That boundary condition is the same as that of
\eq{
\mathscr{A} : \mspace{20mu} q(0) = q_0, \mspace{20mu} q(T) = q_T,
\mspace{20mu} \dot{q}(0) = \dot{q}_0, \mspace{20mu} \dot{q}(T) = \dot{q}_T
}
which was proposed in ref.~\cite{hh}. The quantum action now takes the form
\begin{eqnarray}
S[\tilde{q}] &=& \int_0^T dt \mspace{5mu} \biggl( \frac{1}{2} \dot{\tilde{q}}^2 
- \frac{m^2}{2} \tilde{q}^2
- \frac{\alpha ^2}{2} \ddot{\tilde{q}}^2 \biggr) \nonumber \\
&=& - \frac{1}{2} \int_0^T dt \mspace{5mu} \tilde{q} \biggl( D^2 + m^2 + \alpha ^2 D^4 \biggr) \tilde{q}
+ \frac{1}{2} \biggl[ \tilde{q}\dot{\tilde{q}} - \alpha ^2 \dot{\tilde{q}}\ddot{\tilde{q}}
+ \alpha ^2 \tilde{q}\dddot{\tilde{q}} \biggr] _0^T \la{Sqtilde} \nonumber \\
& &
\end{eqnarray}
where the (last) boundary term vanishes due to the boundary condition (\re{bcon}).

The boundary term in eq.~(\re{Sqtilde}) also vanishes by another boundary condition,
\eq{
\tilde{\mathscr{A}}' : \mspace{20mu} \tilde{q}(0) = 0, \mspace{20mu} \tilde{q}(T) = 0,
\mspace{20mu} \ddot{\tilde{q}}(0) = 0, \mspace{20mu} \ddot{\tilde{q}}(T) = 0 \la{bcon2}
}

As a result, the action (\re{Sqtilde}) takes the Gaussian form, which is quite appropriate for a path
integral quantization with the Gaussian functional
\eq{
- \frac{1}{2} \int_0^T dt \mspace{5mu} \tilde{q} \biggl( D^2 + m^2 + \alpha ^2 D^4 \biggr) \tilde{q}
}
Let us now compute the {\it spectrum} of the operator $D^2 + m^2 + \alpha ^2 D^4$. For this purpose, 
we need to find the solutions $u_k$ to the eigenvalue equation
\eq{
\bigl( D^2 + m^2 + \alpha ^2 D^4 \bigr) u_k (t) = k u_k (t)
}
with the eigenvalues $k$. A general solution is
\begin{eqnarray}
u_k (t) &=& A_1 \cos\bigl( \omega _+ t \bigr) + A_2 \sin\bigl( \omega _+ t \bigr)
+ A_3 \cos\bigl( \omega _- t \bigr) + A_4 \sin\bigl( \omega _- t \bigr) \nonumber \\
& & \omega _\pm = \sqrt{ \frac{1 \mp \sqrt{1 - 4\alpha ^2 (m^2 - k)} }{2\alpha ^2} } \la{uk}
\end{eqnarray}
where $A_1,A_2,A_3,A_4$ is the constants of integration. The function $\tilde{q}$ can be expanded
in terms of $u_k$,
\eq{
\tilde{q} = \int dk \mspace{5mu} u_k (t)
}
The spectrum of $k$ is now  determined by appying the physical boundary conditions (\re{bcon}) or
(\re{bcon2}) to $u_k$ in the form of eq.~(\re{uk}). Applying the boundary condition (\re{bcon}) 
at $t=0$ yields
\eq{
\tilde{q}(0) = A_1 + A_3 = 0 , \mspace{50mu} \dot{\tilde{q}}(0) = A_2 \omega _+ + A_4 \omega _- = 0
}
The boundary condition ({\re{bcon}) at $t=T$ then takes the form
\begin{eqnarray}
\tilde{q}(T) &=& A_1 \cos\bigl( \omega _+ T \bigr) + A_2 \sin\bigl( \omega _+ T \bigr) 
- A_1 \cos\bigl( \omega _- T \bigr) \nonumber \\
& & - A_2 \frac{\omega _+}{\omega _-} \sin\bigl( \omega _- T \bigr) = 0 \nonumber \\
\dot{\tilde{q}}(T) &=& - A_1 \omega _+ \sin\bigl( \omega _+ T \bigr) + A_2 \omega _+ 
\cos\bigl( \omega _+ T \bigr) \nonumber \\
& & + A_1 \omega _- \sin\bigl( \omega _- T \bigr) - A_2 \omega _+ \cos\bigl( \omega _- T \bigr) 
= 0 \nonumber 
\end{eqnarray}
In particular, the determinant of the matrix on the left side of this equation,
\begin{align}
& \text{det}
\begin{pmatrix}
\omega _- \biggl[ \cos\bigl( \omega _+ T \bigr) - \cos\bigl( \omega _- T \bigr) \biggr] &
\omega _- \sin\bigl( \omega _+ T \bigr) - \omega _+ \sin\bigl( \omega _- T \bigr) \\
- \omega _+ \sin\bigl( \omega _+ T \bigr) + \omega _- \sin\bigl( \omega _+ T \bigr) &
\omega _+ \biggl[ \cos\bigl( \omega _+ T \bigr) - \cos\bigl( \omega _- T \bigr) \biggr]
\end{pmatrix} \nonumber \\
=& \omega _+ \omega _- \biggl[ \cos\bigl( \omega _+ T \bigr) - \cos\bigl( \omega _- T \bigr) 
\biggr] ^2 \nonumber \\
& + \omega _+ \omega _- \biggl[ \sin ^2 \bigl( \omega _+ T \bigr) + \sin ^2 \bigl( \omega _- T \bigr) 
\biggr]
- ( \omega _+^2 + \omega _-^2 ) \sin\bigl( \omega _+ T \bigr) \sin\bigl( \omega _- T \bigr) \nonumber \\
=& 2 \omega _+ \omega _- \biggl[ 1 - \cos\bigl( \omega _+ T \bigr) \cos\bigl( \omega _- T \bigr) 
\biggr] - ( \omega _+^2 + \omega _-^2 )
\sin\bigl( \omega _+ T \bigr) \sin\bigl( \omega _- T \bigr)
\end{align}
must vanish. We find
\begin{eqnarray}
2 \omega _+ \omega _- \biggl[ 1 - \cos\bigl( \omega _+ T \bigr) \cos\bigl( \omega _- T \bigr) \biggr]
&=& ( \omega _+^2 + \omega _-^2 ) \sin\bigl( \omega _+ T \bigr) \sin\bigl( \omega _- T \bigr) 
\nonumber \\
\frac{ 2\sqrt{m^2-k} }{\alpha} \biggl[ 1 - \cos\bigl( \omega _+ T \bigr) \cos\bigl( \omega _- T \bigr) 
\biggr]
&=& \frac{1}{\alpha ^2} \sin\bigl( \omega _+ T \bigr) \sin\bigl( \omega _- T \bigr) \nonumber \\
1 - \cos\bigl( \omega _+ T \bigr) \cos\bigl( \omega _- T \bigr)
&=& \frac{1}{\alpha \sqrt{m^2-k} }
\sin\bigl( \omega _+ T \bigr) \sin\bigl( \omega _- T \bigr) \nonumber \\
& &
\end{eqnarray}
where $\omega_{\pm}(k)$ ar given by eq.~(\re{uk}). Apparently, there is no simple solution here.

When employing the boundary conditions (\re{bcon2}) with eq.~(\re{uk}) on $u_k$, the boundary 
condition in $t=0$ yields
\eq{
\tilde{q}(0) = A_1 + A_3 = 0 , \mspace{50mu} \ddot{\tilde{q}}(0) = - A_1 
\omega _+^2 - A_3 \omega _-^2=0
}
so that we find $A_1=A_3=0$ when $\omega _+ \neq \omega _-$. Now the boundary condition at $t=T$ reads
\begin{eqnarray}
\tilde{q}(T) &=& A_2 \sin\bigl( \omega _+ T \bigr) + A_4 \sin\bigl( \omega _- T \bigr) = 0 \nonumber \\
\ddot{\tilde{q}}(T) &=& - A_2 \omega _+^2 \sin\bigl( \omega _+ T \bigr) - A_4 \omega _-^2 \sin\bigl( 
\omega _- T \bigr) = 0
\end{eqnarray}
To get a nontrivial solution, the correspending determinant must vanish, which yields the condition
\eq{
( \omega _+^2 - \omega _-^2 ) \sin\bigl( \omega _+ T \bigr) \sin\bigl( \omega _- T \bigr) = 0
}
Since $\omega _+ \neq \omega _-$, we find
\eq{
\sin\bigl( \omega _+ T \bigr) = 0 \mspace{50mu} \text{or} \mspace{50mu} \sin\bigl( \omega _- T \bigr) 
= 0
}
It means
\eq{
\omega _+ = \frac{n \pi}{T} \mspace{50mu} \text{or} \mspace{50mu} \omega _- = \frac{n\pi}{T}
\mspace{50mu} \bigl( \text{where } n~~ \text{is an integer} \bigr)
}
and $\omega _\pm$ are the solutions to the equation
\eq{
x^2 + m^2 + \alpha ^2 x^4 = k
}
Therefore, the spectrum of $k$ with the boundary condition $\tilde{\mathscr{A}}'$ has the simple 
form
\eq{
k = \biggl( \frac{n\pi}{T} \biggr) ^2 + m^2 + \alpha ^2 \biggl( \frac{n\pi}{T} \biggr) ^4
}

\section{Canonical quantization and instabilities}

In this section we recall about istabilities and ghosts in the quantum PU oscillator
\cite{wood}.
The most straightward way is based on identifying the energy rasing and lowering 
operators \cite{wood}. The classical solution (\re{qcl}) can be rewritten to the form
\begin{eqnarray}
q(t) &=& \frac{1}{2}(A_{+} - i B_{+}) e^{i \lambda _{+} t } + \frac{1}{2}(A_{+} + i B_{+}) 
e^{-i \lambda _{+} t } \nonumber \\
& & + \frac{1}{2}(A_{-} - i B_{-}) e^{i \lambda _{-} t } + \frac{1}{2}(A_{-} + i B_{-}) 
e^{-i \lambda _{+} t }
\end{eqnarray}
Since the $\lambda_-$ modes have negative energy, the lowering operator must be proportional to the
$(A_- -iB_-)$ amplitude. Similarly, since the $\lambda_+$ modes have negative energy, the 
raising  operator must be proportional to the $(A_+ +iB_+)$ amplitude, ie.
\begin{eqnarray}
\alpha_{\pm} &\sim& A_{\pm} \pm i B_{\pm} \nonumber \\ 
&\sim& \frac{\lambda_{\pm}}{2}(1 \pm \sqrt{1-4 \alpha^2 m^2}) Q_{1} \pm iP_{1} 
\mp \frac{i}{2}(1 \mp \sqrt{1-4 \alpha^2 m^2}) - \lambda_{\pm}P_{2}\nonumber \\
& &
\end{eqnarray}
where we have used 
\eq{
A_{\pm} = \frac{\ddot{q}_{0}+\lambda_{\mp}^2 q_{0}}{\lambda_{\mp}^2 - \lambda_{\pm}^2}
}
and
\eq{
B_{\pm} = \frac{\dddot{q}_{0}+\lambda_{\mp}^2\dot{q}_{0}}{\lambda_{\pm}(\lambda_{\mp}^2
-\lambda_{\pm}^2)}
}
as well as~\footnote{The canonical variables were calculated at the initial time value
because the operators in Schrodinger picture do not depend upon time.}
\eq{
Q_{1} = q_{0}
}
\eq{
Q_{2} = \dot{q}_{0}
}
\eq{
P_{1} = \dot{q}_{0} + \alpha^2 \dddot{q}_{0}
}
\eq{
P_{2} = -\alpha^2 \ddot{q}_{0}
}
It is now straightfoward to derive the commutation relations,
\eq{
[\alpha_{\pm},\alpha_{\pm}^{\dagger}]=1
}

The next step depends upon physical interpretation \cite{wood}. 

(I) The `empty' (or `ground') state may be defined by the condition 
\eq{
\alpha_{+} \ket{\bar{\Omega}} = \alpha_{-}^{\dagger} \ket{\bar{\Omega}} = 0
}
Then the `empty' state wave function  $\bar{\Omega}(Q_{1},Q_{2})$ 
(in the $Q$-representation, with $P=-i\partial/\partial Q$) reads
\eq{
\bar{\Omega}(Q_{1},Q_{2}) = N \exp \left[ - \frac{\sqrt{1-4\alpha^2m^2}}{2(\lambda_{-}-\lambda_{+})}
(\lambda_{+}\lambda_{-}Q_{1}^2-Q_{2}^2)-im\alpha Q_{1}Q_{2} \right]
}
and is {\it infinite or not normalizable}, because the size of the wave function gets bigger with 
the 
increase of $Q_{2}$, so that the integral over the whole space diverges. 

In addition, when the eigenstate $\ket{\bar{N}_{+},\bar{N}_{-}}$ with the eigenvalues 
$\bar{N}=(\bar{N}_{+},\bar{N}_{-})$ is defined by
\eq{
\ket{\bar{N}_{+},\bar{N}_{-}}
= \frac{a_{+}^{\dagger}}{\sqrt{N_{+}!}}\frac{a_{-}}{\sqrt{N_{-}!}}\ket{\bar{\Omega}}
}
the norm of the $(0,1)$ state is given by
\begin{eqnarray}
<0,\bar{1}|0,\bar{1}> &=& \bra{\bar{\Omega}} \alpha_{-}^{\dagger}\alpha_{-} \ket{\bar{\Omega}} 
\nonumber \\
&=& \bra{\bar{\Omega}} (-1+\alpha_{-}\alpha_{-}^{\dagger}) \ket{\bar{\Omega}} \nonumber \\
&=& -<\bar{\Omega}|\bar{\Omega}> \nonumber \\
&=& -1
\end{eqnarray}
which is a ghost. The non-normalizable quantum `states' are physically unacceptable, so the 
interpretation (I) should be dismissed \cite{wood}.

(II) It is, however, possible to treat all particles (with positive or negative energy) as the 
truly ones by defining the `empty' state $\Omega$ differently, namely, as
\eq{
\alpha_{\pm} \ket{\Omega} = 0
}
In this interpretation the negative energy can arbitrarily decrease and the Hamiltian is 
unbounded from below. The `empty' state solution $\Omega(Q_{1},Q_{2})$ in the $Q$ representation 
is now given by
\eq{
\Omega(Q_{1},Q_{2}) = N \exp \left[ - \frac{\sqrt{1-4\alpha^2m^2}}{2(\lambda_{-}+\lambda_{+})}
(\lambda_{+}\lambda_{-}Q_{1}^2+Q_{2}^2)+im\alpha Q_{1}Q_{2} \right]
}
and is apparently {\it finite or normalizable}, because the first term in the exponential is 
negative.

The eigenstate $\ket{\bar{N}_{+},\bar{N}_{-}}$ of the eigenvalues 
$\bar{N}=(\bar{N}_{+},\bar{N}_{-})$ 
is now given by
\eq{
\ket{N_{+},N_{-}}
= \frac{a_{+}^{\dagger}}{\sqrt{N_{+}!}}\frac{a_{-}^{\dagger}}{\sqrt{N_{-}!}}\ket{\Omega}
}
while the norm  of the $(0,1)$ state is
\begin{eqnarray}
<0,1|0,1> &=& \bra{\Omega} \alpha_{-}\alpha_{-}^{\dagger} \ket{\Omega} \nonumber \\
&=& \bra{\Omega} (1-\alpha_{-}^{\dagger}\alpha_{-}) \ket{\Omega} \nonumber \\
&=& <\Omega|\Omega> \nonumber \\
&=& 1
\end{eqnarray}
ie. it is not a ghost. 

In the correct physical interpretation (II) the correspondence principle 
between the classical and quantum states is preserved, but the system has indefinite energy. When
interactions are switched on, mixing the negative and positive energy states would lead to 
instabilities in the classical theory, and the exponentially growing and decaying states in 
quantum theory \cite{tomb,nest}. Excluding the negative energy states would lead to the loss of 
unitarity \cite{hh}.

\section{Path integral quantization and Forman theorem}

The idea of ref.~\cite{hh} is to define the quantum theory of the PU oscillator as the 
{\it Euclidean} path integral and then Wick rotate it back to Minkowski case. It makes 
sense since the Euclidean
action of the PU oscillator --- see eq.~(\re{epos}) below --- is positively definite. It can 
also
make the difference to the canonical quantization and the Ostrogradski method (Sec.~2) when 
one integrates over the path {\it only}, but not over its derivatives.  

Let us first recall some basic facts about a path integral in QFT, according to the standard 
textbooks in Quantum Field Theory -- see, for example, ref.~\cite{ryder}.

The definition of the probability amplitude for a one-dimensional quantum particle by 
Feynman path integral is given by
\eq{
Z(q_{b},t_{b};q_{a},t_{a})
=\int ^{q_{b}}_{q_{a}}\mathscr{D}q \exp{\left[\frac{i}{\hbar}\int^{t_b}_{t_a}dtL \right]}
}
where the integration goes over all paths $q(t)$ between $q_a$ and $q_b$. After Wick rotation
\eq{
t \rightarrow t=  -i \tau
}
the path integral takes the form\footnote{The sign factor in the Wick rotation is chosen to
make the path integral converging.}
\eq{
Z(q_{b},t_{b};q_{a},t_{a})
=\int ^{q_{b}}_{q_{a}}\mathscr{D}q \exp{\left[ - \frac{1}{\hbar}\int^{\tau_b}_{\tau_a}d\tau L_E 
\right]}
}
It is called the Euclidean path integral. In the case of the PU oscillator the Euclidean 
path integral is {\it Gaussian}. Let us recall some basic properties of the Gaussian integrals. 

The simplest Gaussian integral reads
\eq{
\int^{\infty}_{-\infty}dx e^{-ax^2} = \sqrt{\frac{\pi}{a}} \qquad a>0
}
It can be easily extended to a quadratic form in the exponential as
\eq{
\int^{\infty}_{-\infty}dx e^{-ax^2-bx} = \sqrt{\frac{\pi}{a}}\exp\left( \frac{b^2}{4a} \right)
}

It can also be easily extended to the case of several variables with the diagonal quadratic form
as
\eq{
\int^{\infty}_{-\infty} [d^{n}x] \exp\left( -\sum_{i=1}^{n}a_{i}x_{i}^{2} \right)
 =  {\frac{1}{\prod_{i=1}^{n} a_{i}^{\frac{1}{2}}}}
}
where we have introduced the normalized measure $[dx]=dx/\sqrt{\pi}$.

By diagonalizing a generic (non-degenerate) quadratic form, one can prove a general
finite-dimensional formula,
\begin{eqnarray}
\int^{\infty}_{\infty} [d^{n}x] \exp\left( -x^{t}Ax -b^{t}x\right) &=& \frac{1}{\prod_{i=1}^{n}
\lambda_{i}}\exp\left(\frac{1}{4}b^{t}A^{-1}b\right) \\ \nonumber
&=& \frac{1}{det A} \exp\left( \frac{1}{4}b^{t}A^{-1}b\right)
\end{eqnarray}

Finally, when formally sending the number of integrations to infinity, one gets the 
Gaussian path integral,
\begin{eqnarray}
& & \int ^{q_{b}}_{q_{a}}\mathscr{D}q \exp{\left[- \int^{t_b}_{t_a} dt ( q(t)F(D)q(t) + q(t)J(t) ) \right]} \nonumber \\
&=& \frac{1}{\sqrt{DetF(D)}} \exp\left[ - \frac{1}{4} \int^{t_b}_{t_a} dt J(t) F^{-1}(D) J(t)
\right]
\end{eqnarray}
where $DetF(D)$ is now the functional determinant.

A generic functional determinant diverges since it is defined as the product of all the 
eigenvalues in the spectrum of a differential operator. Therefore, one needs a regularization.
It is most convenient to use the zeta function regularization in our case --- see, for example,
ref.~\cite{zeta} for a comprehensive account. The Riemann zeta function is defined by
\eq{
\zeta (s) = \sum^{\infty}_{n=1} \frac{1}{n^s}
}
in the convergence area of the series. It is then expanded for ${\rm Re}(s)>1$ by 
analytic continuation. It is often useful to employ an integral representation of the
 zeta function in the form
\eq{
\zeta (s) = \frac{1}{\Gamma (s)}\int^{\infty}_{0} dt \, t^{s-1} \sum^{\infty}_{n=1}e^{-nt}
 \la{zin}
}
where the (Euler) gamma function has  been introduced,
\eq{
\Gamma (s) = \int^{\infty}_{0} dt \, t^{s-1}e^{-t}
}

Equation (\re{zin}) allows one to define the zeta function for an elliptic operator $L$ as
\eq{
\zeta(s|L)=\frac{1}{\Gamma (s)}\int^{\infty}_{0} dt \, t^{s-1} \mathrm{tr} e^{-tL}
}
where $\mathrm{tr} e^{-tL}$ is given by
\begin{eqnarray}
\mathrm{tr} e^{-tL} &=&
\mathrm{tr} \left(
\begin{array}{ccccc}
e^{-\lambda_{1}t} & & & &  \\
&e^{-\lambda_{2}t} & & &  \\
& & \ddots & &  \\
& & & &  \\
\end{array}
\right) \nonumber \\
&=& \sum^{\infty}_{n=1} e^{-\lambda_{n} t}
\end{eqnarray}
in terms of the positive eigenvalues $\lambda_{n}$ of $L$. One easily finds
\begin{eqnarray}
\zeta(s|L) &=& \sum^{\infty}_{n=1} \frac{1}{\lambda^{s}_{n}} \nonumber \\
&=& \sum^{\infty}_{n=0} e^{-s \ln \lambda_{n}}
\end{eqnarray}
Differentiating both sides of this equation with respect to $s$ at $s=0$, one finds
\begin{eqnarray}
\left. \frac{d \zeta (s|L)}{ds} \right| _{s=0} &=& -\sum^{\infty}_{n=1} \ln \lambda_{n} 
\nonumber \\
&=& -\ln \prod ^{\infty}_{n=1} \lambda_{n} \nonumber \\
&=& -\ln Det L
\end{eqnarray}
so that the functional determinant of an elliptic operator $L$ is given by
\eq{
Det L = e^{-{\zeta}^\prime|_{s=0}}
}
The zeta function regularization of the right hand side of this equation is 
\begin{eqnarray}
\ln Det \frac{L(\epsilon)}{\mu^2} &=& - \frac{1}{\epsilon} 
\zeta \left( \epsilon \left|\frac{L}{\mu ^2} \right. \right) \nonumber \\
&=& - \frac{1}{\Gamma (\epsilon + 1)}
\int^{\infty}_{0} dt \, t^{\epsilon -1} \mathrm{tr} e^{-t \frac{L}{\mu^2}} \nonumber \\
&=& -\frac{1}{\epsilon}\sum^{\infty}_{n=1} 
\frac{1}{\left(\frac{\lambda_{n}}{\mu ^2} \right)^\epsilon} \nonumber \\
&=& - \frac{\mu^{2 \epsilon}}{\epsilon} \sum^{\infty}_{n=1} \frac{1}{\lambda ^{\epsilon}_{n}} 
\nonumber \\
&=& - \frac{\mu^{2 \epsilon}}{\epsilon} \zeta(\epsilon|L) \nonumber \\
&=& - \frac{1}{\epsilon}(1+\epsilon \ln \mu^2)(\zeta(0|L)+\epsilon \zeta'(0|L)) + O(\epsilon^2)
\nonumber \\
&=& - \frac{1}{\epsilon}\zeta(0|L)-\zeta'(0|L)-\ln \mu^2 \zeta(0|L) + O(\epsilon^2) \la{zetare}
\end{eqnarray}
where we have introduced the regularization parameter $\epsilon$ and the dimension parameter
$\mu$.

The zeta-function renormalization amounts to deleting the first term in eq.~(\re{zetare}), since
it UV-diverges in the limit $\epsilon \rightarrow 0$, as well as the third term since it 
IR-diverges in the limit $\mu\to 0$.

To put equation (\re{zetare}) into a more explicit form, without resorting to the spectrum of 
the differential  operator, it is convenient to use Forman's theorem 
\cite{forman}:~\footnote{Forman theorem is an extension of the Gel'fand-Yaglom theorem.}

Let $K_{\mathscr{A}}$ and $\Bar{K}_{\Bar{\mathscr{A}}}$ are the differential operators defined by
\begin{equation}
\left\{
\begin{array}{c}
K = P_{0} (\tau )\frac{d^{n}}{d\tau ^{n}} + O(\frac{d^{n-1}}{d\tau ^{n-1}}) \\
\Bar{K} = P_{0} (\tau )\frac{d^{n}}{d\tau ^{n}} + O(\frac{d^{n-1}}{d\tau ^{n-1}})
\end{array}
\right.
\end{equation}
over the domain $[0,T]$.  Consider a linear differential equation
\eq{
Kh(\tau)=0
}
with a boundary condition
\eq{
\mathscr{A} : M
\left( 
\begin{array}{c}
h(0) \\
h^{(1)}(0) \\
\vdots \\
h^{(n-1)}(0)
\end{array}
\right)
+N
\left( 
\begin{array}{c}
h(0) \\
h^{(1)}(0) \\
\vdots \\
h^{(n-1)}(0)
\end{array}
\right)
=0
}
and take the boundary condition $\Bar{\mathscr{A}}$ to be {\it smoothly} 
connected to $\mathscr{A}$. The time evolution operator $Y_{K}(\tau)$ is introduced as
\eq{
\left( 
\begin{array}{c}
h(\tau) \\
\vdots \\
h^{(n-1)}(\tau)
\end{array}
\right)
=
Y_{K}(\tau)
\left( 
\begin{array}{c}
h(0) \\
\vdots \\
h^{(n-1)}(0)
\end{array}
\right)
}
so that the boundary condition can be written to 
\eq{
\left(M+NY_{K}(T) \right)
\left( 
\begin{array}{c}
h(0) \\
\vdots \\
h^{(n-1)}(0)
\end{array}
\right)
=0
}

The Forman theorem is given by the statement:
\eq{
\frac{DetK_{\mathscr{A}}}{Det\Bar{K}_{\Bar{\mathscr{A}}}}
=\frac{det\left(M+NY_{K}(T) \right)}{det\left(\Bar{M}+\Bar{N}Y_{\Bar{K}}(T) \right)}
}

This theorem is effective for finding the functional determinant of the operator $K$ with 
unknown spectrum by connecting it to the one with a simple spectrum via changing the 
boundary conditions.

\section{Path Integral of PU Oscillator}

The Euclidean path integral of the PU oscillator over a domain $[0,T]$ was calculated in 
refs.~\cite{hh,zer,pol}. Here we confirm the results of ref.~\cite{zer} by our calculation.   

The path integral of PU oscillator with the action
\eq{
S_{PU} = \int_{0}^{T} dt \left( \frac{1}{2}\dot{q}(t)^2 - \frac{m^2}{2}q(t)^2 -
 \frac{\alpha^2}{2}\ddot{q}(t)^2 \right)
}
after the Wick rotation $(t \rightarrow it$) takes the form
\eq{
Z(q_{T},T;q_{0},0) = 
\int ^{q_{T}}_{q_{0}}\mathscr{D}q \exp{(-S_{E})}
}
where the Euclidean PU action is given by
\eq{
S_{E}
= \int_{0}^{T} dt \left( \frac{1}{2}\dot{q}(t)^2 + \frac{m^2}{2}q(t)^2 
+ \frac{\alpha^2}{2}\ddot{q}(t)^2 \right) \la{epos}
}
This $S_{E}$ is {\it positively definite}, so that the Euclidean path integral is well 
defined.

Since our discussion of the classical theory (Sec.~4), the integral trajectory is a sum of 
a classical trajectory $q_{cl}$ and quantum fluctuations $\hat{q}$,  $q = q_{cl} + \hat{q}$.
Accordingly, the action can be also written down as a sum,
\eq{
S_{E}[ q ] =S_{cl} + S [ \hat{q} ]
}
and the path integral of the PU oscillator takes the form
\eq{
Z(q_{T},T;q_{0},0) = 
e^{-S_{cl}} \int ^{0}_{0}\mathscr{D} \hat{q} \exp{(-S[\hat{q}])}
}
where the quantum action $S[\hat{q}]$ is given by
\eq{
S[\hat{q}]
= \frac{1}{2} \int_{0}^{T} dt \, \hat{q} \left( \alpha^2 \frac{d^4}{dt^4} 
- \frac{d^2}{dt^2} + m^2  \right) \hat{q}
}
after integration by parts.

Let us denote the differential operator $\alpha^2 \frac{d^4}{dt^4} - \frac{d^2}{dt^2} + m^2$ 
with the boundary condition $\mathscr{A}$ as $K_{\mathscr{A}}$. Then the path integral can be
written down in the form
\eq{
Z(q_{T},T;q_{0},0) = 
e^{-S_{cl}} \int ^{0}_{0}\mathscr{D} \hat{q} 
\exp{ \left(-\frac{1}{2} \int_{0}^{T} dt \, \hat{q} K_{\mathscr{A}} \hat{q} \right)}
}
The path integral of the PU oscillator is Gaussian and, therefore, can be computed along the lines
of Sec.~7 as 
\eq{
Z(q_{T},T;q_{0},0) = \frac{N}{\sqrt{Det K_{\mathscr{A}}}}\exp{(-S_{cl})}
}
where $N$ is the normalization constant. The classical part $S_{cl}$ was found in ref.~\cite{hh},
and it is quite involved. The functional determinant is the key part of a quantum propagator of 
PU oscillator, which is of primary physical interest. It can be computed by the use of Forman
theorem (Sec.~7). 

First, one calculates the time evolution operator $Y_{K}$. It is given by
\eq{
Y_{K}(t) =
\left(
\begin{array}{cccc}
u_{1}(t) &u_{2}(t) &u_{3}(t) &u_{4}(t)  \\
\dot{u_{1}}(t) &\dot{u_{2}}(t) &\dot{u_{3}}(t) &\dot{u_{4}}(t)  \\
\ddot{u_{1}}(t) &\ddot{u_{2}}(t) &\ddot{u_{3}}(t) &\ddot{u_{4}}(t)  \\
\dddot{u_{1}}(t) &\dddot{u_{2}}(t) &\dddot{u_{3}}(t) &\dddot{u_{4}}(t) 
\end{array}
\right)
}
where 
\eq{
Ku_{i}(t)=0 \: (i=1,\dots,4)
}
and the inital condition is
\eq{
Y_{k}(0)=1
}
The operator $K_{\bar{\mathscr{A}}}$ is equal to $K_{\mathscr{A}}$, so they have $Y_{K}(t)$ 
is common.

By solving the equation $Ku_{i}=0$ for $u_{i}$ with
\eq{
K = \alpha ^2 \frac{d^4}{dt^4} - \frac{d^2}{dt^2} + m^2
}
one gets its general solution in the form
\eq{
u_{i}(t) = A_{i}\sinh(\lambda_{+}t) + 
B_{i}\cosh(\lambda_{+}t) + C_{i}\sinh(\lambda_{-}t) + D_{i}\cosh(\lambda_{-}t)
}

The boundary condition $Y_{K}(0)=1$ amounts to the relations
\begin{eqnarray}
B_{i} + D_{i} = \delta_{1i} \\
\lambda_{+} A_{i} + \lambda_{-} C_{i} = \delta_{2i} \\
\lambda_{+}^2 B_{i} + \lambda_{-}^2 D_{i} = \delta_{3i} \\
\lambda_{+}^3 A_{i} + \lambda_{-}^3 C_{i} = \delta_{4i}
\end{eqnarray}
Therefore,  the solutions are
\begin{eqnarray}
u_{1} &=& \frac{\lambda_{-}^2}{\lambda_{-}^2 - \lambda_{+}^2}\cosh(\lambda_{+}t)
 + \frac{\lambda_{+}^2}{\lambda_{+}^2 - \lambda_{-}^2}\cosh(\lambda_{-}t) \\
u_{2} &=& \frac{\lambda_{-}^2}{\lambda_{+}(\lambda_{-}^2 - \lambda_{+}^2)}\sinh(\lambda_{+}t)
 + \frac{\lambda_{+}^2}{\lambda_{2}(\lambda_{+}^2 - \lambda_{-}^2)}\sinh(\lambda_{-}t) \\
u_{3} &=& - \frac{1}{\lambda_{-}^2 - \lambda_{+}^2}\cosh(\lambda_{+}t)
 - \frac{1}{\lambda_{+}^2 - \lambda_{-}^2}\cosh(\lambda_{-}t) \\
u_{4} &=& - \frac{1}{\lambda_{+}(\lambda_{-}^2 - \lambda_{+}^2)}\sinh(\lambda_{+}t)
 + \frac{1}{\lambda_{2}(\lambda_{+}^2 - \lambda_{-}^2)}\sinh(\lambda_{-}t)
\end{eqnarray}

Next, one writes down the boundary conditions $\mathscr{A}$ and $\bar{\mathscr{A}}$ in
terms of the matrices $M$ and $N$ appearing in the Forman theorem. The boundary condition 
$\mathscr{A}$ is 
\eq{
\mathscr{A} \: : \: \hat{q}(0)=0  \: , \: 
\hat{q}(T) = 0 \: , \: \dot{\hat{q}}(0) = 0 \: , \: \dot{\hat{q}}(T)=0
}
so that its matrices $M$ and $N$ are given by
\eq{
M=
\left(
\begin{array}{cccc}
1 &0 &0 &0  \\
0 &1 &0 &0  \\
0 &0 &0 &0  \\
0 &0 &0 &0 
\end{array}
\right)
}
and
\eq{
N=
\left(
\begin{array}{cccc}
0 &0 &0 &0  \\
0 &0 &0 &0  \\
1 &0 &0 &0  \\
0 &1 &0 &0 
\end{array}
\right)
}

In the same way, the boundary condition $\bar{\mathscr{A}}$ is 
\eq{
\bar{\mathscr{A}} \: : \: \hat{q}(0)=0  \: , \: \hat{q}(T) = 0 \: , 
\: \ddot{\hat{q}}(0) = 0 \: , \: \ddot{\hat{q}}(T)=0
}
so that its matrices $M$ and $N$ are given by
\eq{
\bar{M}=
\left(
\begin{array}{cccc}
1 &0 &0 &0  \\
0 &0 &-1 &0  \\
0 &0 &0 &0  \\
0 &0 &0 &0 
\end{array}
\right)
}
and
\eq{
\bar{N}=
\left(
\begin{array}{cccc}
0 &0 &0 &0  \\
0 &0 &0 &0  \\
1 &0 &0 &0  \\
0 &0 &1 &0 
\end{array}
\right)
}

Having found $M$, $N$ and $Y_K$, as well as $\bar{M}$, $\bar{N}$ and $Y_{\bar{K}}$,
we calculate
\begin{align}
& det(M+NY_{K}(T)) \nonumber \\
&= \frac{\alpha^3}{m}\left[ \frac{1}{1+2m\alpha}\sinh^2 
\left(\frac{\sqrt{1+2m\alpha}}{2\alpha}T \right)
- \frac{1}{1-2m\alpha}\sinh^2 \left(\frac{\sqrt{1-2m\alpha}}{2\alpha} T \right) \right]
\end{align}
and
\begin{align}
& det(\bar{M}+\bar{N}Y_{\bar{K}}(T)) \nonumber \\ 
&= \frac{\alpha}{m} \left[ \sinh^2
\left(\frac{\sqrt{1+2m\alpha}}{2\alpha}T \right)
- \sinh^2 \left(\frac{\sqrt{1-2m\alpha}}{2\alpha}T \right) \right]
\end{align}

A calculation of $Det K_{\bar{\mathscr{A}}}$ goes along the standard lines \cite{hh,zer,pol}, 
\begin{eqnarray}
Det K_{\bar{\mathscr{A}}} &=& \prod_{n=1}^{\infty} k_{n} \nonumber \\ 
&=& \prod_{n=1}^{\infty} \left( \alpha^2 \left(\frac{n\pi}{T}\right)^4 +
 \left(\frac{n\pi}{T}\right)^2 +m^2 \right)  \\
&=& \frac{\alpha}{mT^2}\left[ \sinh^2 \left(\frac{\sqrt{1+2m\alpha}}{2\alpha}T \right)
- \sinh^2 \left(\frac{\sqrt{1-2m\alpha}}{2\alpha} T \right) \right] \nonumber
\end{eqnarray}
By using the Forman formula, we get the final answer
\begin{eqnarray}
Det K_{\mathscr{A}}
&=& \frac{det\left(M+NY_{K}(T) \right)}{det\left(\Bar{M}+\Bar{N}Y_{\Bar{K}}(T) \right)}
Det K_{\bar{\mathscr{A}}} \nonumber \\
&=&  \frac{\alpha^3}{mT^2} 
\left[ (1+2\alpha m)^{-1} \sinh^2 \left(\frac{\sqrt{1+2m\alpha}}{2\alpha}T \right) \right. 
 \\
& & \left. \qquad \qquad \qquad - (1-2\alpha m)^{-1} \sinh^2 \left(\frac{\sqrt{1-2m\alpha}}{2
\alpha} T \right) \right] \nonumber
\end{eqnarray}
in full agreement with ref.~\cite{zer} in its last (v2) version. In the large $T\gg 1$ limit 
one finds 
\begin{align}
& Det K_{\mathscr{A}} =  \\
&  \frac{\alpha}{m}\left[ \frac{1}{1+2\alpha m} \frac{\exp \left(\sqrt{1+2m\alpha}
\frac{T}{\alpha}\right)}{\left(\frac{2T}{\alpha} \right)^2}
- \frac{1}{1-2\alpha m} \frac{\exp \left(\sqrt{1-2m\alpha}\frac{T}{\alpha}
\right)}{\left(\frac{2T}{\alpha} \right)^2} \right] \nonumber
\end{align}
and in the small $T$ limit one gets
\eq{
Det K_{\mathscr{A}} \approx \frac{T^2}{12} +{\cal O}(T^4)  
}

The transition amplitude (or the quantum Euclidean propagator) of PU oscillator 
is given by 
\eq{
<q_T,\dot{q}_T;\tau=T | q_0,\dot{q}_0;\tau=0 >
= \sqrt{ \frac{2\pi}{Det  K_{\mathscr{A}}} } \exp\left( -S_E[q_{cl}]\right) 
}
The classical Euclidean action $S_E[q_{cl}]$ was calculated in Appendix of ref.~\cite{hh}.
It is finite for large $T\gg 1$ and behaves like $\frac{1}{2T}$ for small $T\ll 1$. Hence, 
the transition amplitude is exponentially suppressed both for small and large $T$, ie. it 
is normalizable, and the Euclidean path integral is well defined indeed.

\section{Conclusion} 

The procedure of calculating Euclidean transition probabilities (for observables) in the 
quantum PU theory was outlined in ref.~\cite{hh}. The probabilities in the Minkowski space 
can be obtained by analytic continuation. It is, therefore, possible to make physical 
sense out of the quantum PU theory. 

In classical PU theory with interactions, even at a very small value of the parameter 
$\alpha>0$, one gets runaway production of states with negative and positive energy. 
However, as was suggested in ref.~\cite{hh}, the Euclidean formulation of the quantum 
theory implicitly imposes certain restrictions that can remove classical instabilities.
The price of removing the instabilities is given by an apparent violation of unitarity 
\cite{hh}. Indeed, integrating over the basic trajectory, and not over its derivatives
in the Euclidean path integral formulation of the quantum PU oscillator given above 
is not in line with the canonical quantization and the Ostrogradski method. By doing it,
one looses some information and, hence, one loses unitarity. As was argued in 
ref.~\cite{hh}, one can, nevertheless, never produce a negative norm state or get a 
negative probability, so that the departure from unitarity may be very small at the low
energies (say, in the present universe), but important at the very high energies (say, in 
the early universe).  Of course, it is debateable whether the `price' of loosing unitarity 
is too high or not. 

Apparently, the $f(R)$ gravity theories are special in the sense that for each of them
there exist the classically equivalent scalar-tensor field theory without higher 
derivatives, under the physical stability conditions. Still, as the quantum field 
theories, they may be different. Hence, it may be possible to quantize $f(R)$ gravity without
loosing unitarity. Figuring  out the details is still a challenge.

\end{document}